\documentclass[conference]{IEEEtran}
\IEEEoverridecommandlockouts
\usepackage{cite}
\usepackage{amsmath,amssymb,amsfonts}
\usepackage{bm}
\usepackage{textcomp}
\usepackage{xcolor}
\usepackage{tikz}
\usetikzlibrary{shapes, arrows, positioning, calc, fit}
\usepackage{amsmath}
\usepackage[T1]{fontenc}
\usepackage{graphicx}
\usepackage{cite}
\usepackage{algpseudocode}
\usepackage{amsmath,amssymb,amsfonts}
\usepackage{amsthm}
\usepackage{textcomp}
\usepackage{xcolor}
\usepackage{subcaption}
\usepackage{upgreek}
\usepackage{booktabs}
\newcommand{\mycomment}[1]{}
\usepackage{balance}
\usepackage[hidelinks]{hyperref}
\def\BibTeX{{\rm B\kern-.05em{\sc i\kern-.025em b}\kern-.08em
    T\kern-.1667em\lower.7ex\hbox{E}\kern-.125emX}}

\makeatletter
\newcommand{\linebreakand}{%
  \end{@IEEEauthorhalign}
  \hfill\mbox{}\par
  \mbox{}\hfill\begin{@IEEEauthorhalign}
}

\usepackage[ruled,vlined,linesnumbered]{algorithm2e}

\begin{document}

    \title{Assessing the Impact of Vaccination on Rotavirus Transmission Dynamics Using Bayesian Inference\\
\thanks{This research was funded in whole, or in part, by the Wellcome Trust [226691/Z/22/Z]. For the purpose of Open Access, the author has applied a CC BY public copyright license to any Author Accepted Manuscript version arising from this submission. CR was funded by the Wellcome CAMO-Net UK grant: 226691/Z/22/Z; JM was funded by a Research Studentship jointly funded by the EPSRC Centre for Doctoral Training in Distributed Algorithms EP/S023445/1 and SM were funded by EPSRC through the Big Hypotheses under Grant EP/R018537/1\\
\bf{\footnotesize \textcopyright 2025 IEEE. Personal use of this material is permitted.
  Permission from IEEE must be obtained for all other uses, in any current or future
  media, including reprinting/republishing this material for advertising or promotional
  purposes, creating new collective works, for resale or redistribution to servers or
  lists, or reuse of any copyrighted component of this work in other works.}}
}

\author{\IEEEauthorblockN{Conor Rosato\IEEEauthorrefmark{1}, Joshua Murphy\IEEEauthorrefmark{2}, Simon Maskell\IEEEauthorrefmark{2}
and
John Harris\IEEEauthorrefmark{3}}\\
\IEEEauthorblockA{\IEEEauthorrefmark{1} Department of Pharmacology and Therapeutics, University of Liverpool, United Kingdom \\
\IEEEauthorrefmark{2} Department of Electrical Engineering and Electronics, University of Liverpool, United Kingdom\\
\IEEEauthorrefmark{3} United Kingdom Health Security Agency (UKHSA), United Kingdom\\
Email: \{cmrosa, joshua.murphy, smaskell\}@liverpool.ac.uk, john.p.harris@ukhsa.gov.uk}}


\maketitle

\begin{abstract}

The introduction of the rotavirus vaccine in the United Kingdom (UK) in 2013 led to a noticeable decline in laboratory reports in subsequent years. To assess the impact of vaccination on rotavirus transmissibility we calibrated a stochastic compartmental epidemiological model using Sequential Monte Carlo (SMC) methods. Our analysis focuses on estimating the time-varying transmissibility parameter and documenting its evolution before and after vaccine rollout. We observe distinct periods of increasing and decreasing transmissibility, reflecting the dynamic response of rotavirus spread to immunization efforts. These findings improve our understanding of vaccination-driven shifts in disease transmission and provide a quantitative framework for evaluating long-term epidemiological trends.

\end{abstract}

\begin{IEEEkeywords}
Bayesian inference; parameter estimation; statistical signal processing; Sequential Monte Carlo; disease survelliance; time-varying parameters.
\end{IEEEkeywords}

\section{Introduction}\label{sec:introduction}

Rotavirus is a leading cause of acute gastroenteritis in young children resulting in significant morbidity and mortality worldwide. Before the widespread introduction of rotavirus vaccines, infections were responsible for a substantial number of hospitalizations and deaths, particularly in infants and young children. The World Health Organization (WHO) has emphasized the importance of vaccination in reducing the global disease burden by recommending routine immunization as an effective public health intervention \cite{rota_who}. The introduction of the rotavirus vaccine in the United Kingdom (UK) in 2013 led to a decline in general practice visits \cite{thomas2017impact} and a reduction in hospital admissions due to rotavirus-related gastroenteritis \cite{marlow2015assessing}. Despite this success, the long-term impact of vaccination on the dynamics of rotavirus transmission remains an area of active research. The interaction between vaccination coverage, waning immunity, and seasonal transmission patterns presents a difficult epidemiological challenge \cite{ai2020disease}.

A simple approach to evaluating the impact of rotavirus vaccination compared expected incidence trends with observed case reductions. For example interrupted time-series models outlined shifts in trends pre- and post- vaccine \cite{hungerford2018rotavirus, otieno2020impact, biggart2018lack}. Metrics such as incidence rate ratios (IRRs) allow for the quantification of changes in infection rates by offering a straightforward comparison \cite{thomas2017impact, verberk2021impact}. Additionally, change point detection methods have been applied to identify critical time periods where transmission dynamics shift \cite{hungerford2019reduction}.

The Susceptible, Infected and Recovered disease model \cite{kermack1927contribution} and its extensions have been widely used when modelling rotavirus transmission \cite{omondi2015modeling, weidemann2014modelling, bilcke2015quantifying, asare2020modeling, geard2022model, kraay2022predicting, lee2024prediction}. By calibrating these models to empirical data using Bayesian methods, estimates can be made of key epidemiological parameters such as the basic reproduction number and growth rate \cite{storvik2023sequential, temfack2024review, del2015sequential}. See \cite{weidemann2014modelling, bilcke2015quantifying, roman2015using} for examples of Bayesian calibration of rotavirus transmission. Incorporating time-varying parameters is crucial when calibrating these models with multiple years worth of data as disease transmission dynamics evolve over time due to factors like vaccination and seasonality \cite{bouman2024bayesian}. Static parameters may not accurately reflect these evolving conditions resulting in biased estimates. Time-varying parameters allow the model to adapt to these changes. Bayesian inference can be performed using Markov Chain Monte Carlo (MCMC) \cite{brooks1998markov} and Sequential Monte Carlo (SMC) methods \cite{andrieu2010particle}. The particle filter (PF) \cite{arulampalam2002tutorial} is a SMC method that is well-suited for modeling disease dynamics that contain time-varying parameters within the Susceptible, Infected, and Recovered framework \cite{endo2019introduction, dureau2013capturing, papageorgiou2024novel}. By sequentially updating parameter estimates as new data become available, PFs effectively capture changes in disease transmission. Bayesian methods provide a natural framework for incorporating prior knowledge and handling uncertainty in parameter estimates which is especially useful for assessing the impact of vaccination programs over time.

This study contributes to the understanding of rotavirus transmission dynamics by providing a detailed, data-driven analysis of time-varying parameters before and after the introduction of the rotavirus vaccine in the UK. We provide a framework for calibrating a stochastic compartmental model with SMC methods and offer new insights into the impact of vaccination on disease spread. More specifically, we highlight key periods of transmissibility fluctuations and demonstrate the effectiveness of this approach for assessing vaccination interventions on long-term trends.

The structure of this paper is organized as follows: Sections~\ref{sec:particle_filter} and \ref{sec:pMCMC} define the PF and PMCMC algorithms, respectively. Section \ref{sec:Examples} outlines the epidemiological model used in the analysis as well as the data used and numerical results. Concluding remarks are presented in Section~\ref{sec:conclusions}.

\section{Particle Filter}\label{sec:particle_filter}
\noindent We define a state transition model
\begin{equation}\label{xt}
\mathbf{x}_{t} | \mathbf{x}_{t-1} \sim f(\mathbf{x}_{t} | \mathbf{x}_{t-1}, \bm{\theta}),
\end{equation}
\noindent and an observation equation 
\begin{equation}\label{yt}
\mathbf{y}_{t} | \mathbf{x}_{t} \sim g(\mathbf{y}_{t} | \mathbf{x}_{t}, \bm{\theta}),
\end{equation}
\noindent where $\bm{\theta}$ represents the parameters with $D$ dimensions.

The goal of the PF is to estimate the posterior distribution of the latent states $p(\mathbf{x}_{1:t} | \mathbf{y}_{1:t})$ given the sequence of observations $\mathbf{y}_{1:t}$. The PF approximates the posterior distribution recursively over time using a set of weighted particles. A particle is a weighted sample from the distribution of latent states. The PF works in two main steps: prediction and update.

At time $t = 0$, a set of $N_x$ particles, $\mathbf{x}_0^{(j)}$ for $j = 1, \dots, N_x$, is drawn from an initial distribution $q(\mathbf{x}_0)$, typically chosen to be the prior distribution $p(\mathbf{x}_0)$. Each particle is assigned an equal weight:
\begin{equation}
    \mathbf{w}_0^{(j)} = \frac{1}{N_x}.
\end{equation}
These particles and their associated weights form the initial approximation to the posterior distribution:
\begin{equation}
    p(\mathbf{x}_0 | \mathbf{y}_0) \approx \sum_{j=1}^{N_x} \mathbf{w}_0^{(j)} \delta(\mathbf{x}_0 - \mathbf{x}_0^{(j)}),
\end{equation}
where $\delta(\cdot)$ is the Dirac delta function representing the particle locations.

At each subsequent time step $t$, particles are propagated based on the state transition model. The new particles $\mathbf{x}_t^{(j)}$ are drawn from a proposal distribution, typically the predicted state based on the previous particle:
\begin{equation}
    \mathbf{x}_t^{(j)} \sim q(\cdot | \mathbf{x}_{t-1}^{(j)}).
\end{equation}
The prediction step incorporates the state dynamics to propagate the particles forward. Each particle is then associated with a new weight $\mathbf{w}_t^{(j)}$ that reflects how well the particle's state matches the observed data.

The weight of each particle is updated based on the likelihood of the observation $\mathbf{y}_t$ given the particle’s state $\mathbf{x}_t^{(j)}$. Using Bayes' theorem, the new weight is computed as:
\begin{equation}
    \mathbf{w}_t^{(j)} = \mathbf{w}_{t-1}^{(j)} \cdot \frac{p(\mathbf{y}_t | \mathbf{x}_t^{(j)})}{q(\mathbf{x}_t^{(j)} | \mathbf{x}_{t-1}^{(j)}, \mathbf{y}_t)},\label{eq:LL}
\end{equation}
where \(p(\mathbf{y}_t | \mathbf{x}_t^{(j)})\) is the likelihood of the observation, $\mathbf{y}_t$ given particle state $\mathbf{x}_t^{(j)}$ and $q(\mathbf{x}_t^{(j)} | \mathbf{x}_{t-1}^{(j)}, \mathbf{y}_t)$ is the proposal distribution.

When $t = 1$, the weights are updated as:
\begin{equation}
    \mathbf{w}_1^{(j)} = \frac{p(\mathbf{y}_1 | \mathbf{x}_1^{(j)}) p(\mathbf{x}_1^{(j)})}{q(\mathbf{x}_1^{(j)} | \mathbf{y}_1)}.
\end{equation}

To ensure that the weights sum to 1, the weights are normalized:
\begin{equation}
    \tilde{\mathbf{w}}_t^{(j)} = \frac{\mathbf{w}_t^{(j)}}{\sum_{j=1}^{N_x} \mathbf{w}_t^{(j)}}.
\end{equation}
This normalization step is essential for ensuring the weights can be used to approximate expectations w.r.t. $p(\mathbf{x}_{1:t}|\mathbf{y}_{1:t})$.

If the weights of the particles become too concentrated (i.e., a small number of particles have significant weights), resampling is performed to avoid degeneracy (i.e., the problem where all but one particle has negligible weight). Resampling selects particles according to their normalized weights and duplicates particles with high weights while discarding those with low weights. The resampling step ensures that the set of particles continues to provide a good approximation to the posterior distribution.

After resampling, the estimate of the posterior distribution is given by the weighted sum of functions of the particles. For example, the expected value of a function $f(\mathbf{x}_t)$ under the posterior can be computed as:
\begin{equation}
    \hat{f}_t = \sum_{j=1}^{N_x} \tilde{\mathbf{w}}_t^{(j)} f(\mathbf{x}_t^{(j)}).
\end{equation}

\section{Particle-Markov Chain Monte Carlo} \label{sec:pMCMC}

Particle Markov Chain Monte Carlo (PMCMC) is an extension of the PF that enables sampling from the posterior distribution $p(\mathbf{x}_{1:t}, \bm{\theta} \mid \mathbf{y}_{1:t})$, including both the latent states $\mathbf{x}_{1:t}$ and the parameters $\bm{\theta}$ of the model. To perform PMCMC, we need to evaluate the likelihood of the observed data, which is typically a challenging task in high-dimensional or non-linear settings. The PF provides a way to approximate this likelihood efficiently, which is then used in conjunction with MCMC methods. The log-likelihood is central to the PMCMC framework. Given a set of observations $\mathbf{y}_{1:t} = \{ \mathbf{y}_1, \dots, \mathbf{y}_t \}$, the log-likelihood function is defined as
\begin{equation}
\log \hat{p}\left(\mathbf{y}_{1:T}|\bm{\theta}\right)\approx\log(\frac{1}{N_x}\sum_{j=1}^N  \mathbf{w}_{t}^{j}), \label{eq:likelihood_pf}
\end{equation}
where \(\mathbf{w}_t^{(j)}\) are the unnormalized importance weights associated with the \(j\)-th particle (as is described in \eqref{eq:LL}). 

In MCMC methods, the accept/reject step is fundamental to ensuring that the chain explores the parameter space correctly and converges to the desired posterior distribution. The accept/reject decision is based on the ratio of the likelihood of the proposed state to the current state. In the context of PMCMC, the log-likelihood is used to guide the sampling of the model parameters $\bm{\theta}$. Given a proposed set of parameters $\bm{\theta}^{*}$ and the corresponding set of latent states $\mathbf{x}_{1:t}^{*}$, the acceptance ratio $a$ for the accept/reject step is computed as
\begin{equation}
    a = \frac{\hat{p}(\mathbf{y}_{1:t}|\bm{\theta}^{*}) p(\bm{\theta}^{*})}{\hat{p}(\mathbf{y}_{1:t}|\bm{\theta}) p(\bm{\theta})},
\end{equation}
where  $\hat{p}(\mathbf{y}_{1:t}|\bm{\theta})$ and $\hat{p}(\mathbf{y}_{1:t} |\bm{\theta}^{*})$ are the likelihoods approximated at the current and proposed parameter values, respectively and $p(\bm{\theta})$ and $p(\bm{\theta}^{*})$ are the prior distributions for the model parameters.

\section{Epidemiological Model}\label{sec:Examples}

The model used in this study is the same as provided in \cite{endo2019introduction} which is adapted from the original study \cite{dureau2013capturing}. The model follows a continuous-time compartmental SEIR framework, which describes the dynamics of disease transmission by categorizing the population into susceptible (\(S\)), exposed (\(E\)), infected (\(I\)), and recovered (\(R\)) compartments. However, for practical implementation and inference, the model is represented in discrete time, allowing for numerical approximation of the differential equations. The transition between compartments is governed by a set of stochastic differential equations, incorporating both deterministic transmission dynamics and stochastic variability in the transmission rate. The cumulative number of new infections (\(Z\)) is updated weekly, resetting at regular intervals to align with observational data. This discrete-time formulation facilitates parameter estimation using p-MCMC while preserving the underlying continuous-time disease dynamics. The full model is presented as:
\begin{align}
\frac{\mathrm{d}S}{\mathrm{d}t} &= - \beta S(t) \frac{I(t)}{N}, \\
\frac{\mathrm{d}E}{\mathrm{d}t} &= \beta S(t) \frac{I(t)}{N} - k E(t), \\
\frac{\mathrm{d}I}{\mathrm{d}t} &= k E(t) - \gamma I(t), \\
\frac{\mathrm{d}R}{\mathrm{d}t} &= \gamma I(t), \\
\frac{\mathrm{d}Z}{\mathrm{d}t} &= k E(t) - Z \delta(t \bmod 7), \\
\frac{\mathrm{d}x}{\mathrm{d}t} &= \sigma \mathrm{dW}, \\
\beta &= e^{x(t)},
\end{align}
where \(Z\) represents the cumulative number of new infections recorded over a seven-day period, and \(x\) is the log-transformed transmissibility. The Dirac delta function, \(\delta(t \bmod 7)\), resets \(Z\) to zero at weekly intervals (\(t \bmod 7 = 0\)). The latent and infectious periods are given by \(\frac{1}{k}\) and \(\frac{1}{\gamma}\), respectively. The transmission rate $\beta$ undergoes Brownian motion in the log scale with fluctuations governed by \(\sigma\). These equations are discretized using fourth-order Runge-Kutta as in \cite{del2015sequential}. We use the same likelihood described in \cite{endo2019introduction}:
\begin{align}
  \mathcal{L}(Z|\mathbf{y}_t, \tau) = \mathcal{N}(\log(\mathbf{y}_t), \log (Z/10), \tau),
\end{align}
which describes the observations as being log-normally distributed. The parameters we are inferring are $\bm{\theta} = \{k, \gamma, \sigma, \tau\}$.

The effective reproduction number, $R_t$, is computed by:
\begin{align}\label{eq:rt}
R_t = R_0S(t),
\end{align}
where $R_0 = \frac{\beta}{N\gamma}$.

The growth rate $r$ of the epidemic can be approximated from the effective reproduction number \eqref{eq:rt}. One common approximation is given by:
\begin{equation}
    r \approx \frac{\gamma (R_t - 1)}{1 + \alpha},
\end{equation}
where $\alpha$ is a factor related to the exposed period defined as $\alpha = \frac{\gamma}{k}$. Substituting $\alpha$ into the equation:
\begin{equation}
r = k \left( \frac{R_t - 1}{1 + \frac{\gamma}{k}} \right).
\end{equation}

Assuming a short latent period, $k \gg \gamma$, it simplifies to:
\begin{equation}
r \approx k (R_t - 1).
\end{equation}

\subsection{Data}\label{sec:Data}

\begin{figure*}
    \centering
    \includegraphics[width=0.7\linewidth]{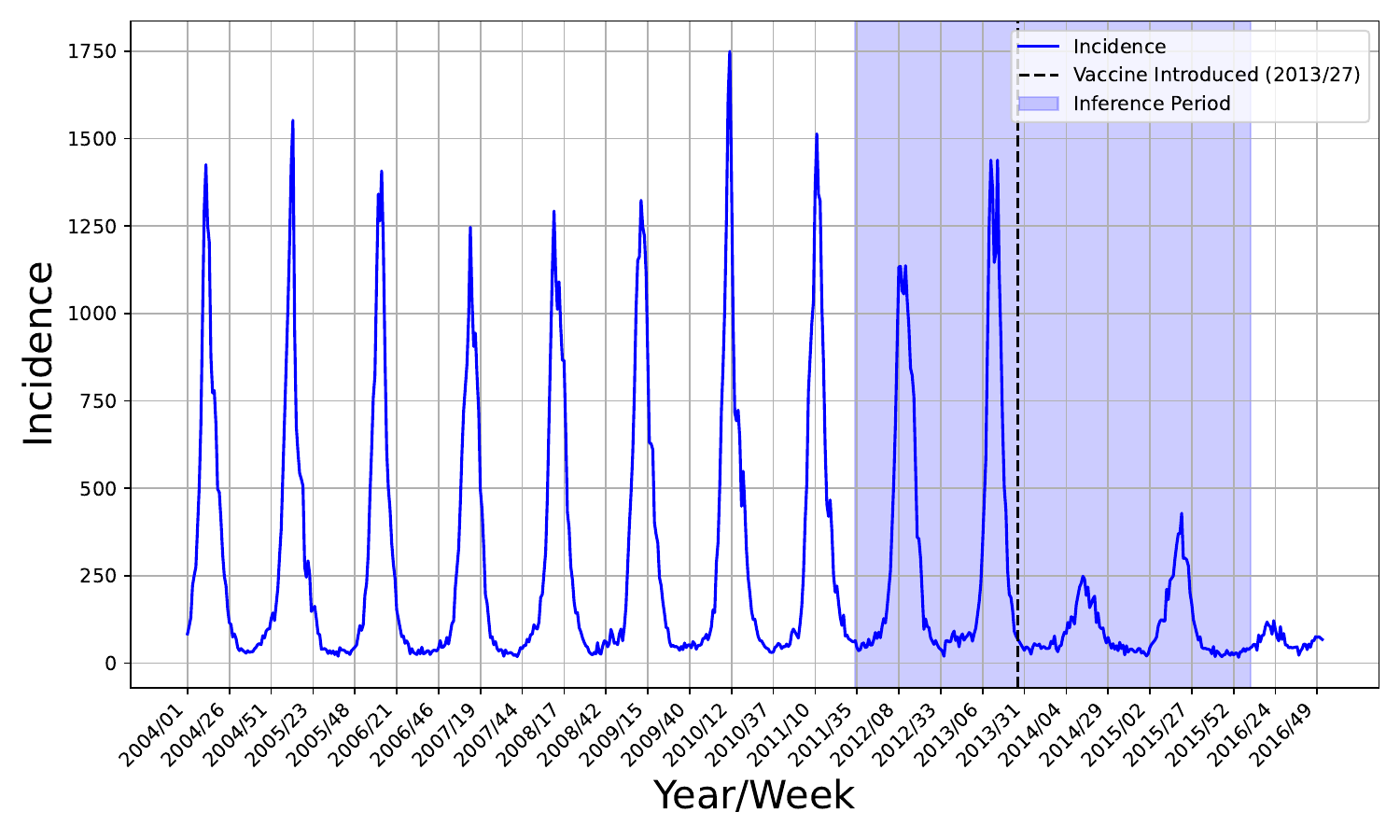}
    \caption{Weekly rotavirus counts in the United Kingdom from 2004-2016. The introduction of the vaccine in 2011 is shown by the black dashed line. The shaded blue section is the portion of the data used in the analysis in Section~\ref{sec:results}.}
    \label{fig:rota_counts}
\end{figure*}

The data used in this paper comprises confirmed cases of rotavirus from 2004 to 2016 and can be seen in Figure~\ref{fig:rota_counts}. The black dashed line is when the vaccine was introduced. The periodicity of confirmed cases is clearly visible, with cases rising during the winter. However, peaks in the years after the vaccine was introduced are remarkably smaller than before. 

\subsection{Results}\label{sec:results}

\begin{figure*}
    \centering
    \subfloat[]{\includegraphics[width=0.48\textwidth]{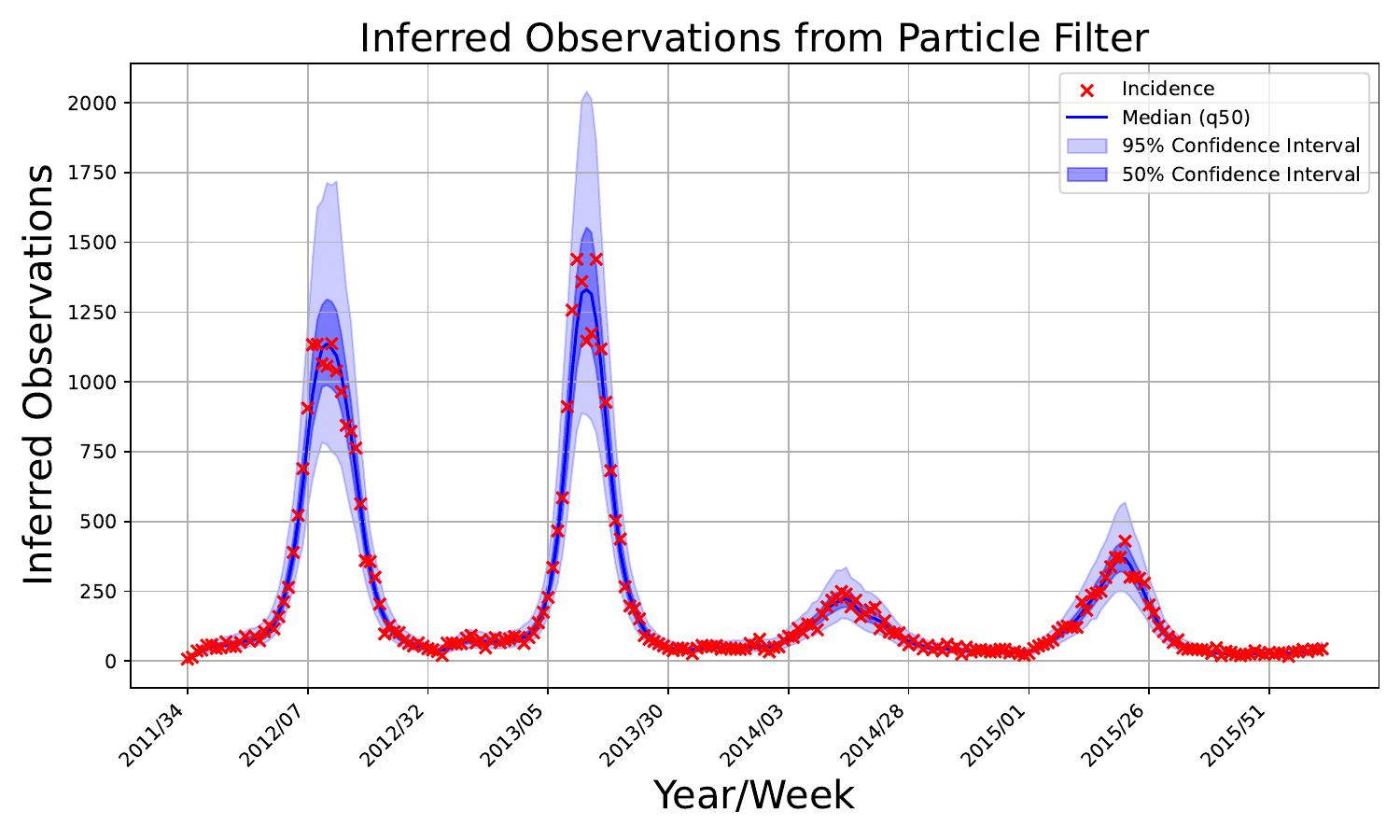}}
    \subfloat[]{\includegraphics[width=0.48\textwidth]{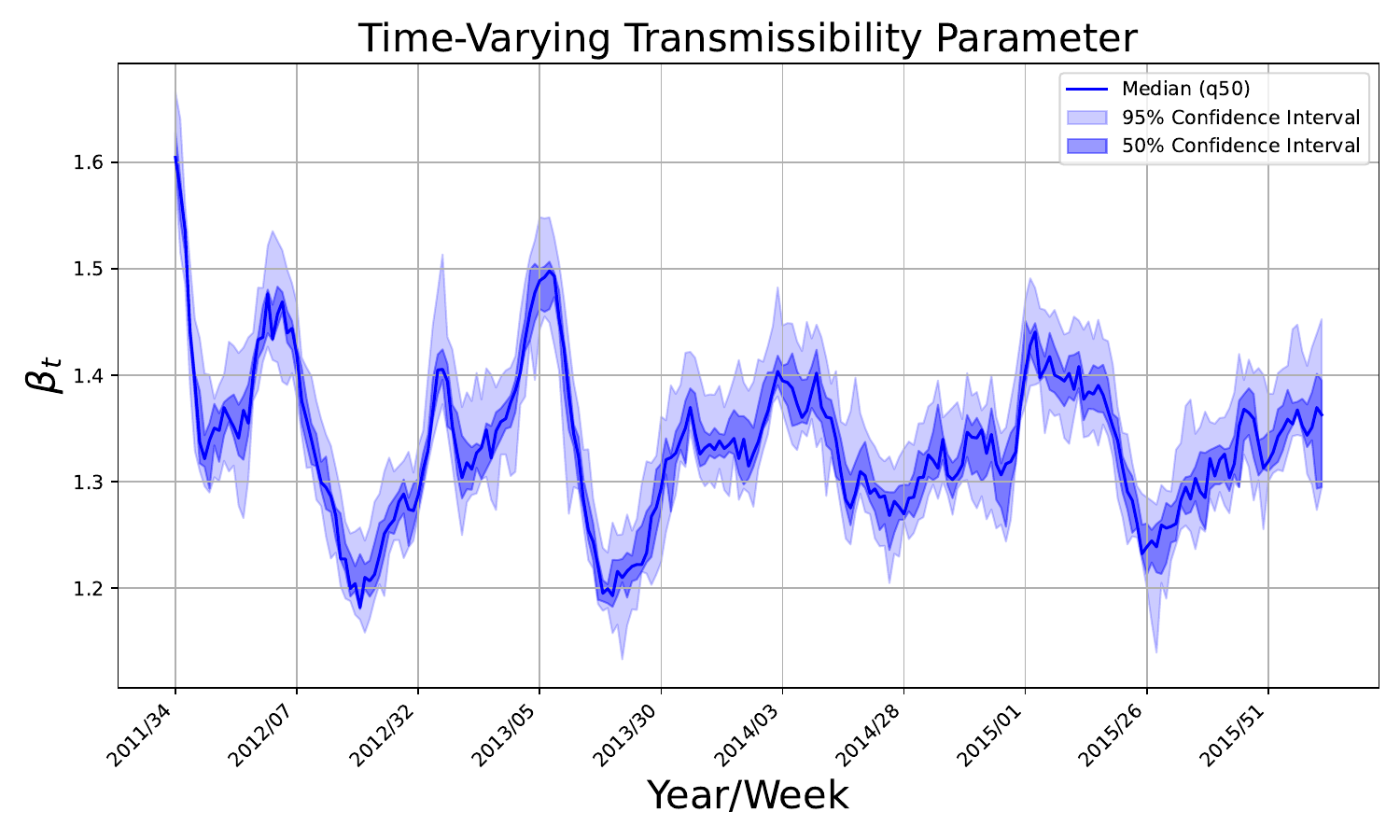}}\\
    \subfloat[]{\includegraphics[width=0.48\textwidth]{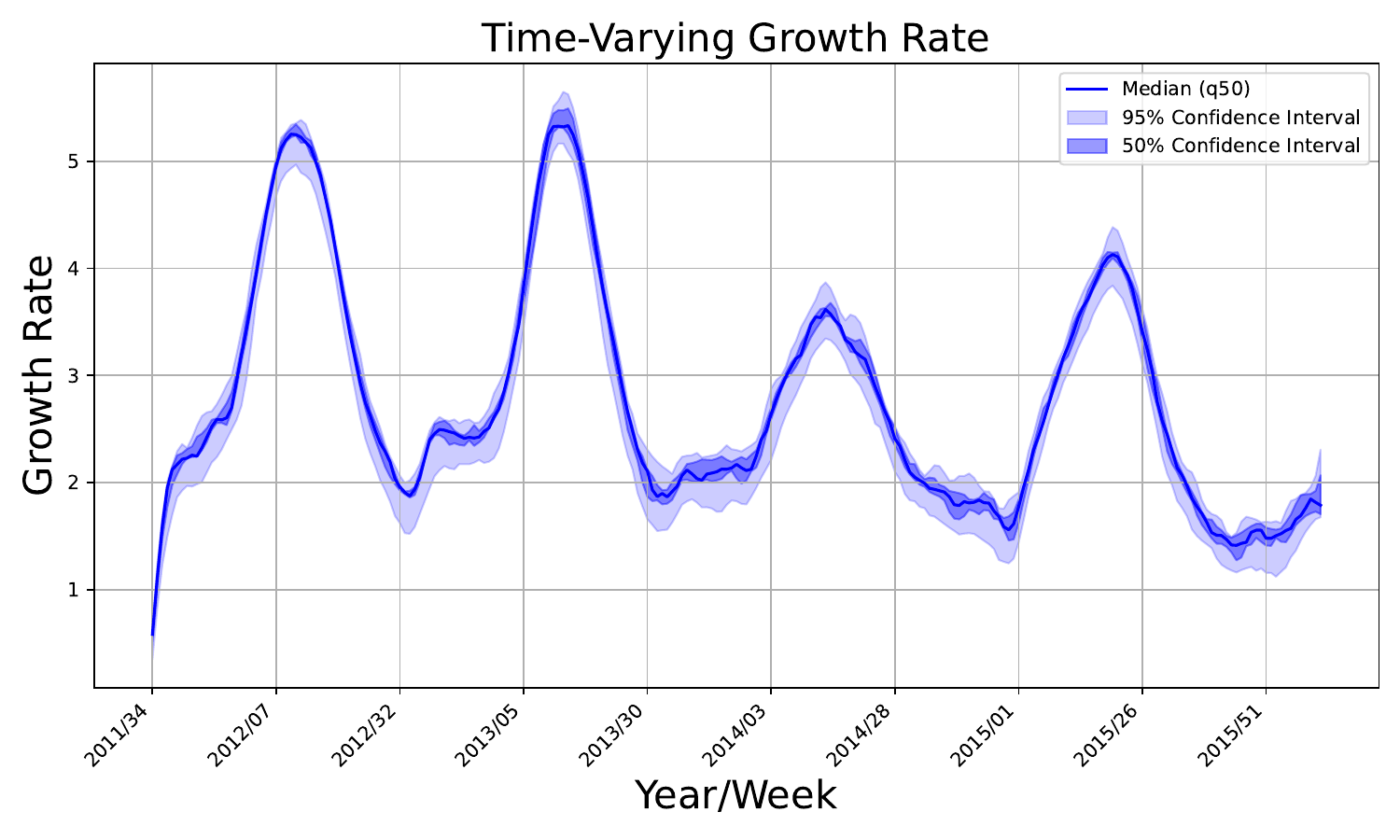}}
    \subfloat[]{\includegraphics[width=0.48\textwidth]{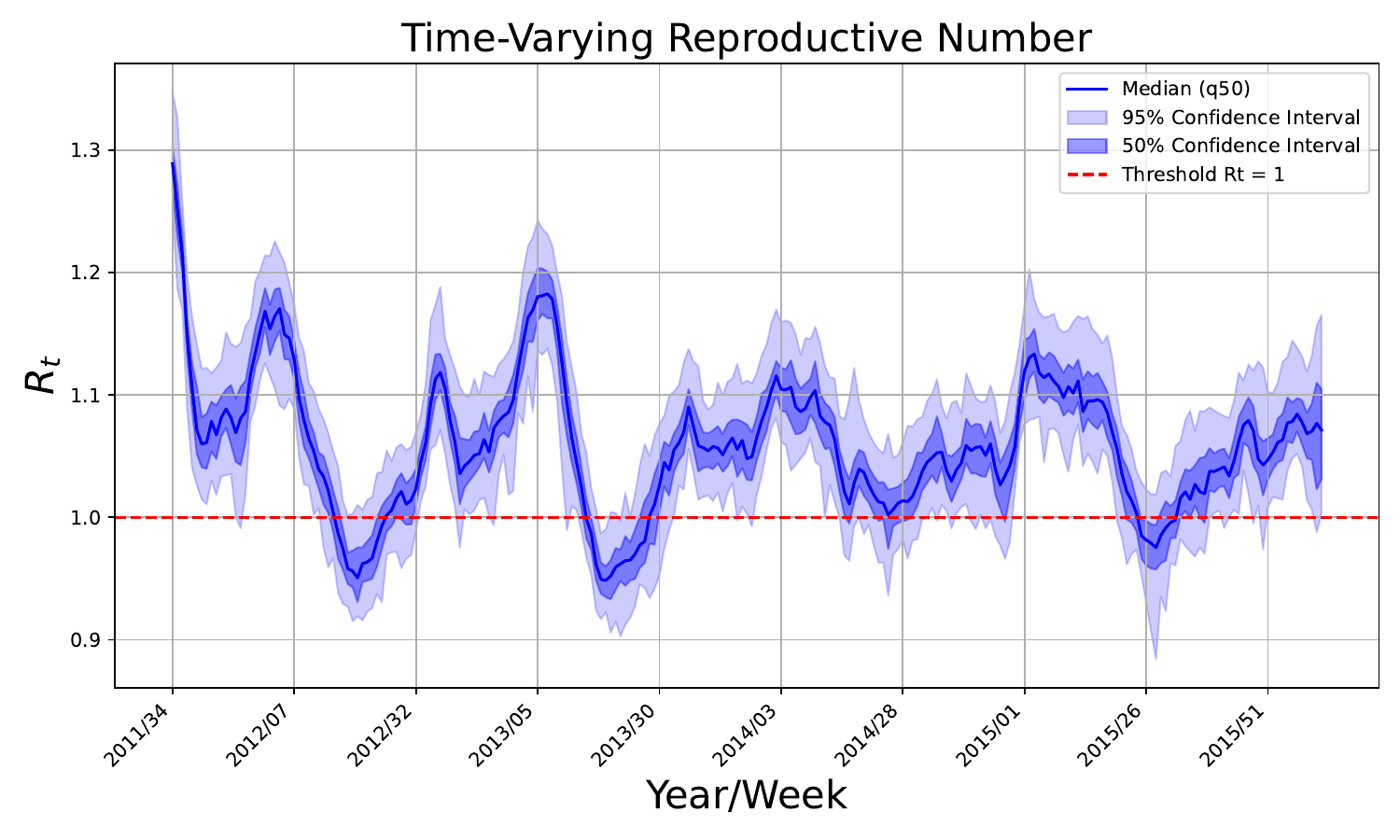}}
    
    \caption{The results produced from the inference process. (a) The posterior predictive distribution from the PF with the true counts represented by red crosses. (b) time-varying transmissibility parameter (c) time-varying growth rate. (d) time-varying reproductive number with the horizontal red line indication a number of 1. In all plots the solid blue line is the median of the sample, while the 50\% and 95\% confidence intervals presented by the shaded regions.}
    \label{fig:results}
\end{figure*}

The analysis performed in this section uses the light blue shaded region of Figure~\ref{fig:rota_counts}. Particle-MCMC was used to fit the epidemiological model in Section~\ref{sec:Examples} to the rotavirus data. The Bayesian inference framework LibBi was used \cite{murray2015bayesian, del2015sequential} and run on a standard 8 core laptop. The sampling process took roughly seven hours. We employed the same adaption methods as described in \cite{endo2019introduction} which helped the efficiency of the sampling process. Figure~\ref{fig:results}(a) shows the inferred observations from the PF closely capture the seasonal trends in rotavirus incidence, reflecting the periodicity observed in the raw data. This form of a posterior predictive distribution serves as a crucial validation of the inference process by demonstrating that the model captures key patterns in the observed data. Figures~\ref{fig:results}(b), (c) and (d) show the time-varying transmutability parameter, growth rate and reproductive number, respectively. These plots show distinct epidemiological patterns. Prior to vaccination, there is a recurrent seasonal trend, peaking around March during the spring months. After the vaccine rollout, these peaks become progressively lower, indicating a sustained reduction in transmission. This pattern is mirrored in the time-varying growth rate, which exhibits a more pronounced seasonal decline post-vaccination, suggesting a weakening of outbreak intensity.

\section{Conclusions and Future Work}\label{sec:conclusions}

This study analyzed the impact of rotavirus vaccination in the UK by estimating the time-varying transmissibility, growth rate, and reproductive number using a stochastic compartmental model calibrated with SMC methods. Our findings highlight clear reductions in transmissibility and outbreak intensity following vaccine introduction, aligning with observed declines in seasonal case peaks. These results reinforce the effectiveness of vaccination in altering disease transmission dynamics and mitigating large seasonal outbreaks.

A potential avenue for future work could sensibly extend this framework by using a hierarchical model that accounts for multiple age groups, as seen in \cite{weidemann2014modelling, bilcke2015quantifying, kraay2022predicting, lee2024prediction}, and incorporate age-structured contact patterns to refine transmission dynamics. A higher dimensional model could potentially need a higher performance PF as seen in \cite{varsi2024general}.

Beyond its epidemiological benefits, rotavirus vaccination has substantial economic implications \cite{marlow2015assessing, thomas2017impact, standaert2023economic, roman2015using}. An active area of research focuses on quantifying the economic benefits of vaccine roll-out, particularly in terms of healthcare cost savings and broader societal impacts. Incorporating a financial aspect to the model presented in this paper would be an interesting direction for future work.

\section*{Acknowledgment}

The authors would like to thank Alexander Phillips for his valuable comments.
\bibliographystyle{ieeetr}
\bibliography{bibliography.bib}


\end{document}